\documentclass[doublecol]{epl2}
\usepackage{units}
\usepackage{amsfonts}
\usepackage{amssymb}
\usepackage{amsmath}
\usepackage{graphicx}
\usepackage{dcolumn}
\usepackage{bm}

\title{Spin transfer torque on magnetic insulators}
\shorttitle{Spin transfer torque on magnetic insulators}

\author{Xingtao Jia\inst{1} \and Kai Liu\inst{1} \and Ke Xia\inst{1} \ \footnote{E-mail: kexia@bnu.edu.cn} and %
Gerrit E. W. Bauer\inst{2,3}}
 \shortauthor{X. Jia \etal}

\institute{
  \inst{1} Department of Physics, Beijing Normal University, Beijing 100875, China\\
  \inst{2} Institute for Materials Research, Tohoku University, Sendai 980-8577,
  Japan\\
  \inst{3} Delft University of Technology, Kavli Institute of NanoScience, 2628 CJ Delft, The Netherlands}

\abstract{Recent experimental and theoretical studies focus on
spin-mediated heat currents at interfaces between normal metals and
magnetic insulators. We resolve conflicting estimates for the order
of magnitude of the spin transfer torque by first-principles
calculations. The spin mixing conductance $G^{\uparrow \downarrow }$
of the interface between silver and the insulating ferrimagnet
Yttrium Iron Garnet (YIG) is dominated by its real part and of the
order of 10$^{14}$ $\unit{\Omega}^{-1}\unit{m}^{-2}$, \textit{i.e.}
close to the value for intermetallic interface, which can be
explained by a local spin model.}
\begin{document}
\maketitle
\section{Introduction}

It has recently been reported that the magnetism of insulators can
be actuated electrically and thermally by normal metal contacts
\cite{Kajiwara2010,Uchida2010}. The material of choice is the
ferrimagnet Y$_{3}$Fe$_{5}$O$_{12}$ (YIG), because of its extremely
small magnetic damping
\cite{Geller1957,Cherepanov1993a,Cherepanov1993b}. The low-lying
excitations of magnetic insulators are spin waves, which carry heat
and angular momentum \cite{Schneider2008}. Existing experiments use
Pt contacts, which by means of the inverse spin Hall effect are
effective spin current detectors \cite{Saitoh06}. Slonczewski
\cite{Slonczewski2010} reports that the thermal spin transfer torque
in magnetic nanopillars \cite{Hatami,Yu} can be much more efficient
than the electrically generated spin torque in metallic structures.

The electrical and thermal injection of spin and heat currents into
insulating magnets is governed by the spin transfer torque at the
metal$|$insulator interface \cite{Xiao2010,Adachi2010}, which is
parameterized by the spin-mixing conductance
$G^{\uparrow\downarrow}=e^{2}\mathrm{Tr}\left(
\mathbf{I}-\mathbf{r}_{\uparrow}^{\dag}\mathbf{r}_{\downarrow}\right)
/h,$ where $\mathbf{I}$ and $\mathbf{r}_{\sigma}$ are the unit
matrix and the matrix of interface reflection coefficients for spin
$\sigma$ spanned by the scattering channels at the Fermi energy of
the metal \cite{Brataas06}. Crude approximations such as a Stoner
model with spin-split conduction bands \cite{Xiao2010} and
parameterized exchange between the itinerant metal electrons and
local moments of the ferromagnet
\cite{Kajiwara2010,Slonczewski2010,Adachi2010} have been used to
estimate $G^{\uparrow\downarrow}$ for YIG interfaces\footnote{The
spin mixing conductance is governed by the reflection coefficients
only and remains finite when the transmission coefficients vanish.
This is not a breach of the scattering theory of transport,since the
incoming and outgoing scattering states are well defined as
propagating states in the metallic contacts.}. Experiments and
initial theoretical estimates found very small spin torques that are
at odds with Slonczewski's predictions \cite{Slonczewski2010}.

Here we report calculations of the spin mixing conductance for the Ag$|$YIG
interface based on realistic electronic structures. Silver is a promising
material \cite{Kimura} for non-local spin current detection \cite{jedema},
which should be more efficient than the inverse spin Hall effect in
nanostructures. We demonstrate that the calculated $G^{\uparrow\downarrow}$
for the Ag$|$YIG interface is much larger than expected from the Stoner model
and better described by local-moment exchange fields.

\section{Free-electron model}

We start with a reference structure consisting of an Ag$|$FI$|$Ag(001)
junction in which the ferromagnetic insulator (FI) is modeled by a spin-split
vacuum barrier, \textit{i.e.,} the free-electron Stoner model. The vacuum
potential is chosen to be spin-split by $0.3$ and $3.0\,%
\operatorname{eV}%
$, whereas the barrier height is adjusted to 0.3, 1.4, 2.6 and $2.85\,%
\operatorname{eV}%
$, respectively. The barrier thickness ($1.2\,%
\operatorname{nm}%
$) is chosen here such that electron transmission is negligible. Table
\ref{Ag_Vac_Ag} lists the corresponding $G^{\uparrow\downarrow}$ of Ag$|$%
FI$|$Ag. Both $%
\operatorname{Re}%
G^{\uparrow\downarrow}$ and $%
\operatorname{Im}%
G^{\uparrow\downarrow}$ decrease with increasing barrier height, as expected
\cite{Xiao2010}. \begin{table}[ptb]
\caption{Spin-dependent and spin mixing conductances of a Ag$|$FI$|$Ag(001)
junction with different barrier heights and spin splitting $\Delta=0.3\ $and
$3.0\ \operatorname{eV}$. The mixing conductances for the (111) orientation
differs by less than 20\%. The Sharvin conductance (G$^{Sh}$) of Ag(001) is
$4.5\times10^{14}\operatorname{\Omega}^{-1}\operatorname{m}^{-2}$.}%
\label{Ag_Vac_Ag}
\begin{center}%
\begin{tabular*}
{8.5cm}[c]{@{\extracolsep{\fill}}lcccc}\hline\hline Barrier &
$G^{\uparrow}$/$G^{sh}$ & $G^{\downarrow}$/$G^{sh}$ & Re$G^{\uparrow
\downarrow}$/$G^{sh}$ & Im$G^{\uparrow\downarrow}$/$G^{sh}$\\\hline
\multicolumn{5}{l}{$\Delta=0.3\operatorname{eV}$}\\
0.3 & 6.3E-5 & 5.1E-6 & 0.009 & -1.1E-1\\
1.4 & 3.3E-8 & 7.1E-9 & 0.003 & -7.4E-2\\
2.6 & 3.5E-9 & 1.3E-9 & 0.001 & -4.0E-2\\
2.85 & 0* & 0 & 0.001 & -5.1E-2\\\hline
\multicolumn{5}{l}{$\Delta=3.0\ \operatorname{eV}$}\\
0.3 & 7.0E-6 & 0 & 0.15 & -0.45\\
1.4 & 7.4E-10 & 0 & 0.08 & -0.35\\
2.6 & 0 & 0 & 0.05 & -0.28\\
2.85 & 0 & 0 & 0.04 & -0.27\\\hline\hline
\multicolumn{5}{l}{{*} 0 means a transmission probability of less than
10$^{-10}$}%
\end{tabular*}
\end{center}
\end{table}

\section{Band structure}

\begin{figure}[ptb]
\includegraphics[width=8.5cm]{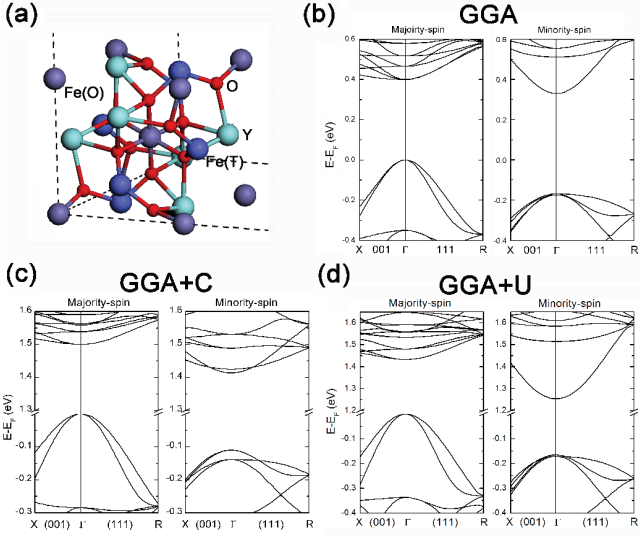}\newline\caption{(a): 1/8 of the cubic
YIG cell; the full structure can be obtained by symmetry operations.
Here, Fe(T) and Fe(O) are Fe atoms at tetragonal and octahedronal
sites, respectively. (b-d): Band structures of YIG with GGA, GGA+C,
and GGA+U method
with band gap of 0.33, 1.25, and $1.25\,\operatorname{eV}$, respectively.}%
\label{yig_band}%
\end{figure}

We calculate the electronic structure of YIG using the tight-binding
linear-muffin-tin-orbital code in the augmented spherical wave approximation
as implemented in the Stuttgart code \cite{stuttgart,stuttgart2,stuttgart3}
using the generalized gradient correction (GGA) to the local density
approximation (LDA). The cubic lattice constant $a=12.2\,%
\operatorname{\text{\AA}}%
$ is chosen 1.6\% smaller than the experimental one \cite{Baettig2008}. We use
136 additional empty spheres (ES) for better space filling and reduced overlap
between neighboring atomic spheres.\textbf{\ }YIG is a ferrimagnetic insulator
with band gap of $2.85\,%
\operatorname{eV}%
$ \cite{Metselaar1974,Wittekoek1975}. Magnetism is carried by majority and
minority spin Fe atoms (tetragonal Fe(T) and octahedral Fe(O) sites in
Fig.\ref{yig_band}(a), respective.) with a net magnetic moment of $5\mu_{B}$
per formula unit \cite{Gilleo1980,Pascard1984,Baettig2008,Rodic1999}. The
magnetic moments are 3.95 and $-4.06\,\mu_{B}$ for majority and minority spin
Fe atoms, respectively. Both Y and O atoms show small positive magnetic
moments of 0.03 and 0.09$\,\mu_{B}$, respectively, while those on the empty
spheres do not exceed 0.007$\,\mu_{B}$. The common problem of
density-functional theory to predict the energy gap of insulators can be
handled by an on-site Coulomb correction (LDA/GGA+U) \cite{lda+u,ching2001} or
a scissor operator (LDA/GGA+C) \cite{lda+c}. Figure \ref{yig_band}(b) is a
plot of the band structure of GGA with a fundamental band gap of $0.33\,%
\operatorname{eV}%
$ between the valence band edge of the majority-spin channel and conductance
band edge of minority-spin channel. The GGA+C method can be used to increase
the band gap depending on the scissor parameters C. A GGA+C band structure
with a band gap of $\backsim$$1.25\,%
\operatorname{eV}%
$ is shown in Fig. \ref{yig_band}(c). The GGA+U method applied to the YIG band
structure using the parameters from Ref. \cite{ching2001,lda+u} leads to the
band structure plotted in \ref{yig_band}(d) with the same energy gap
$\backsim1.25\,%
\operatorname{eV}%
$.

\begin{table}[pth]
\caption{Band gap ($E_{g}$) and effective masses (in unit of $m_{e}$) of the
band structure of YIG at the $\Gamma$ point as calculated by the GGA, GGA+U,
and GGA+C methods. CB and VB denote conductance and valence bands,
respectively.}%
\label{uandc}
\begin{center}%
\begin{tabular*}
{8.5cm}[c]{@{\extracolsep{\fill}}cccccc}\hline\hline
& $E_{g}\left(  \operatorname{eV}\right)  $ &
\multicolumn{2}{c}{Majority-spin} & \multicolumn{2}{c}{Minority-spin}\\
&  & VB & CB & VB & CB\\\hline
GGA & 0.33 & 0.10 & 0.52 & 0.40 & 0.17\\\hline
GGA+U$^{a}$ & 1.25 & 0.13 & 0.60 & 0.37 & 0.19\\
GGA+C$^{b}$ & 1.25 & 0.17 & 1.00 & 0.31 & 0.27\\\hline
GGA+C$^{c}$ & 1.4 & 0.17 & 1.00 & 0.28 & 0.31\\
GGA+C$^{d}$ & 1.4 & 0.18 & 1.46 & 0.25 & 0.25\\\hline\hline
\multicolumn{6}{l}{$^{a}U=3.5\,\operatorname{eV},J=0.8\,\operatorname{eV}$}\\
\multicolumn{6}{l}{$^{b}C(Fe,Y)=6.1\,\operatorname{eV}%
,C(ES)=3.05\,\operatorname{eV}$}\\
\multicolumn{6}{l}{$^{c}C(Fe,Y)=7.2\,\operatorname{eV}$}\\
\multicolumn{6}{l}{$^{d}C(Fe,Y)=7.5\,\operatorname{eV}%
,C(ES)=3.75\,\operatorname{eV}$}%
\end{tabular*}
\end{center}
\end{table}

While a visual comparison of the band structures in Figs. \ref{yig_band}(c)
and (d) assures the equivalence of the two methods, we can assess the
differences quantitatively by comparing the effective masses at the band edges
as shown in Table \ref{uandc}. For band gaps of $\backsim1.25\,%
\operatorname{eV}%
$ the effective mass at the conductance band edge of majority-spin as obtained
by the GGA+U and GGA+C methods differ by up to 67\%. This seems significant,
but the effects on the mixing conductance, which is the quantity of our main
interest here, is small, as discussed in the next section.

\begin{figure}[ptb]
\includegraphics[width=8.0cm]{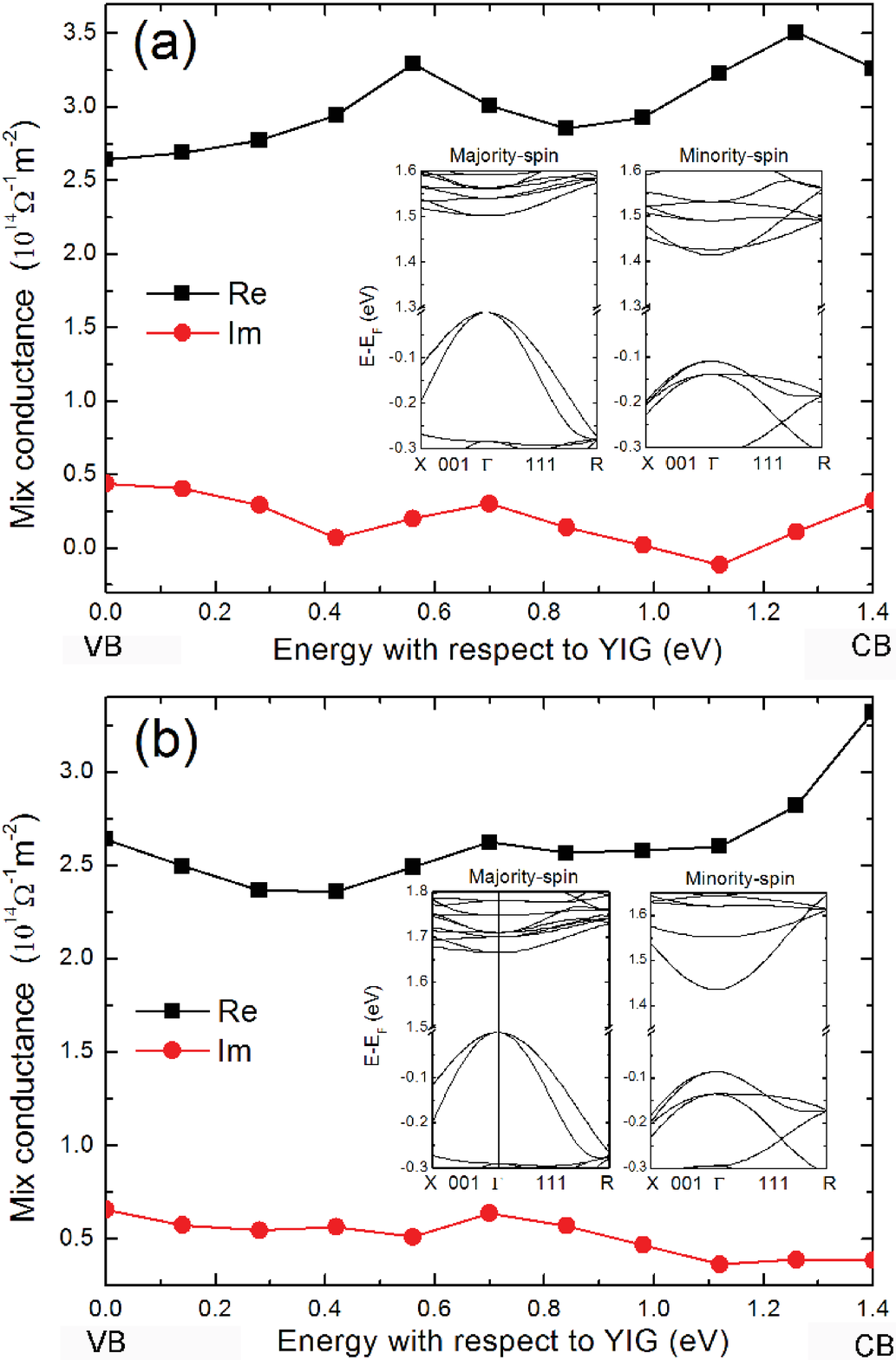}\newline\caption{Effect of band
dispersion and band alignment on the spin mixing conductance of Ag$|$%
YIG$|$Ag(001) with YFe-termination. We use YIG with same band gap of 1.4eV
with different implementations of scissor operator (a)
C(Fe,Y$)=7.2\,\operatorname{eV}$; (b) C(Fe,Y$)=7.5\,\operatorname{eV}$, and
C(ES$)=3.75\,\operatorname{eV}$ to see the effect of band dispersion. We fix
the Fermi energy of Ag while scanning the YIG work function.}%
\label{band_dispersion}%
\end{figure}

\section{Ag$|$YIG interface}

We study the spin mixing conductance in Ag$|$YIG$|$Ag with a 3$\times$3 and
6$\times$6 lateral supercell of fcc Ag to match a cubic YIG unit cell along
the (001) and (111) directions with lattice mismatch $\sim1\%$. Interfaces can
be classified according to their magnetic surface properties into three types
of terminations. For the (001) texture, one cut is terminated by Y as well as
majority and minority spin Fe atoms with compensated magnetic moment
(\textquotedblleft YFe-termination\textquotedblright). Another cut yields only
majority Fe atoms at the interface (\textquotedblleft
Fe-termination\textquotedblright) with total magnetic moment of $7.90\,\mu
_{B}$ per lateral unit cell. The third interface covered by O atoms is
obtained by removing Fe and Y atoms from the YFe-termination. The oxygen layer
is separated from adjacent Fe atoms by only $\sim0.3\,%
\operatorname{\text{\AA}}%
$. Including the latter, the \textquotedblleft O-termination\textquotedblright%
\ also corresponds to a net interface magnetic moment of $7.90\,\mu_{B}$. The
interfaces for the (111) direction can be classified analogously. The
\textquotedblleft YFe-termination\textquotedblright\ cut has now a net
interface magnetic moment of $23.70\,\mu_{B}$. The Fe-termination contains now
minority-spin Fe atoms with net magnetic moment of $-16.24\mu_{B}$, while the
O-terminated surface has the same magnetic moment when including the shallowly
buried Fe layer.

We chose a YIG film of 4 unit cell layers, because its electric conductance
does not exceed $10^{-10}\,e^{2}$/$h$ per unit cell. $G^{\uparrow\downarrow}$
is therefore governed solely by the single Ag$|$YIG interface.

First, we inspect $G^{\uparrow\downarrow}$ of Ag$|$YIG interfaces computed
with and without scissor corrections. We find that the difference of $%
\operatorname{Re}%
G^{\uparrow\downarrow}$ is as small as 21\% when increasing the band gap of
YIG from its GGA value of $0.33\,%
\operatorname{eV}%
$ to a GGA+C band gap of $2.1\,%
\operatorname{eV}%
$ as shown in Table \ref{ldatoc}. We conclude that the precise band gap is a
parameter that hardly affects the $G^{\uparrow\downarrow}$ of the Ag$|$YIG
interface. \begin{table}[pth]
\caption{Spin mixing conductance of Ag$|$YIG(001) with different YIG band gaps
modulated by the GGA+C methods. We pin the Fermi level of Ag at mod-gap of
YIG. $0.33\,\operatorname{eV}$ is the band gap of GGA (without a scissor
operator).}%
\label{ldatoc}
\begin{center}%
\begin{tabular*}
{8.5cm}[c]{@{\extracolsep{\fill}}lcccccc}\hline\hline
$E_{g}(\operatorname{eV})$ & 0.33 & 0.65 & 0.95 & 1.4 & 1.8 & 2.1\\\hline
\multicolumn{7}{l}{$\operatorname{Re}G^{\uparrow\downarrow}$}\\
$(10^{14}\operatorname{\Omega}^{-1}\operatorname{m}^{-2})$ & 3.46 & 3.94 &
3.43 & 3.01 & 2.82 & 2.74\\\hline\hline
\end{tabular*}
\end{center}
\end{table}

By scanning the Fermi energy of Ag (or the YIG work function), we can obtain
information similar to that when varying the band gap. Scanning the Ag Fermi
energy from the valence to conductance band edges for a band gap of $1.4\,%
\operatorname{eV}%
$, we obtain results equivalent to a mid-gap Fermi energy and band gaps
varying from zero to $2.8\,%
\operatorname{eV}%
$, but without changing the details of the band dispersion. In Figure
\ref{band_dispersion} we plot the mixing conductance of Ag$|$YIG(001) with YFe
termination as a function of YIG's work function. Here we consider two kinds
of band dispersions with the same band gap of 1.4$\,%
\operatorname{eV}%
$ obtained by different scissor operator implementations as shown in
Table.\ref{uandc}. We find that the mixing conductance does not depend
sensitively on (i) the YIG work function or interface potential barrier as
well as (ii) the band dispersion when fixing the Fermi energy of Ag in the
middle of the band gap of YIG; the difference in effective mass of
46\thinspace\% causes changes in $%
\operatorname{Re}%
G^{\uparrow\downarrow}$ of only 13\%. These deviations are within the error
bars due to other approximations (see below). We therefore conclude that the
transport properties in the present system are sufficiently well represented
by the scissor operator or on-site Coulomb correction methods for the gap problem.

Besides the band alignment discussed in the previous paragraph, two more
properties are difficult to compute self-consistently for large unit cells,
\textit{viz}. the atomic interface configuration and the ferromagnetic
proximity effect: (i): We determine the distance between Ag$|$YIG by
minimizing ASA overlap while keeping the space filled. We estimate that the
differences in $G^{\uparrow\downarrow}$ for configurations with maximum and
minimum ASA overlap is less than 30\% (ii): We assess the ferromagnetic
proximity effect by using the self-consistent electronic structure of Ag atom
at the Ag$|$Fe interface. We find that the Ag atoms closest to Fe acquire a
magnetic moment of 0.025$\,\mu_{B}$ and the effect is observable up to the 4th
Ag layer. The spin mixing conductance is found to be enhanced by about 10\% in
this system. In the following we disregard such an effect. From various checks
of these and other issues, the magnitude of a possible systematic error in the
mixing conductance is estimated to be $<40\%$.

\begin{figure}[ptb]
\includegraphics[width=8.6cm]{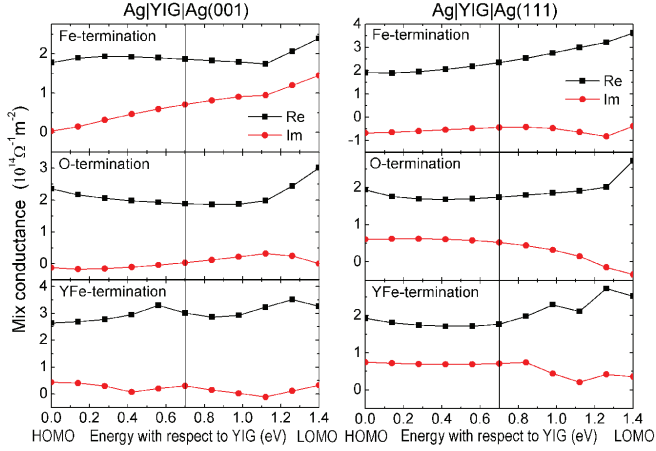}\newline\caption{Spin mixing
conductance of Ag$|$YIG(001) (left) and Ag$|$YIG(111) (right) with a
YIG band gap of $1.4\ \operatorname{eV}$. We fix the Fermi energy of
Ag while scanning
the YIG work function.}%
\label{agyigag_gmix}%
\end{figure}

\section{Results}

Fig. \ref{agyigag_gmix} summarizes our results for $G^{\uparrow\downarrow}$ of
Ag$|$YIG(111) and Ag$|$YIG(001) junction with different YIG
interface-terminations. We find $G^{\uparrow\downarrow}\simeq10^{14}\,%
\operatorname{\Omega }%
^{-1}%
\operatorname{m}%
^{-2}$ for both Ag$|$YIG(111) and Ag$|$YIG(001), with the real part
dominating over the imaginary one, in stark contrast to the Stoner
model (cf. Table 1). $G^{\uparrow\downarrow}$ depends only weakly on
exposing different YIG(111) surface cuts to Ag, which we attribute
to the homogeneous distribution of magnetic atoms. The YFe
termination of YIG(001) is a nearly compensated magnetic interface,
but we still calculate a large spin mixing conductance. Finally, our
results are two orders of magnitude larger than the experimental
value found for Pt$|$YIG(111) \cite{Kajiwara2010}! The difference
between Ag and Pt cannot account for this discrepancy: one would
rather expect a larger $G^{\uparrow\downarrow}$ for Pt because of
its higher conduction electron density.

The difference between the Stoner model and the first-principles calculations
indicate that the spin-transfer torque physics at normal metal interfaces with
YIG is very different from those with transition metals. Spin-transfer is
equivalent to the absorption of a spin current at an interface that is
polarized transversely to the magnetization direction. Magnetism in insulators
is usually described in a local moment model. The physical picture of spin
transfer appropriate for metals, \textit{viz.} the destructive interference of
precessing spins in the ferromagnet, then obviously fails. When the spin
transfer acts locally on the magnetic ions, we expect no difference for the
spin absorbed by a fully ordered interface with a large net magnetic moment or
a compensated one, in which the local moments point in opposite directions, as
is indeed born out of our calculations.

\begin{figure}[ptb]

\includegraphics[width=8.5cm]{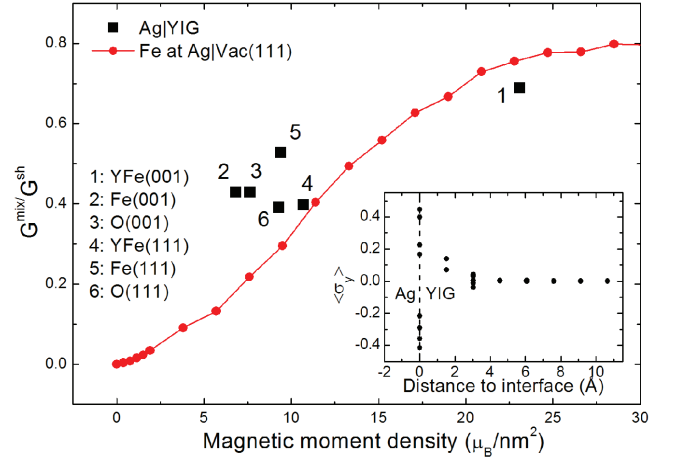}\newline
\caption{$\operatorname{Re}%
G^{\uparrow\downarrow}$ as a function of interface magnetic moment density in
Ag$|$YIG compared with that of Fe atoms at the Ag$|$Vac interface. Insert:
Crystal-plane resolved spin density $\left\langle \sigma_{y}\right\rangle $ in
arbitrary units for Ag$|$YIG(001) with YFe-termination when fully polarized
electrons are injected from the Ag side at mid-gap energy. The maximum
magnetic moment density is 39$\mu_{B}$/$\operatorname{nm}^{2}$ as estimated
from a full monolayer of Fe at the interface, in which the Fe atoms adopt the
Ag structure with magnetic moment of $2.81\mu_{B}$.}%
\label{local_picture}%
\end{figure}

In order to test the local moment paradigm, we consider
non-conducting Ag$|$Vac(4L)$|$Ag(111) junctions. We now sprinkle one
vacuum interface randomly with Fe atoms. At low densities the Fe
atoms are weakly coupled and form local moments. The electronic
structure is generated using the Coherent Potential Approximation
(CPA) for interface disorder. A $10\times10$ lateral supercell with
100 atoms in one principle layer is used to model a magnetic
impurity range from 1\% to 80\%. The high density limit is a
monolayer of Fe atoms in the fcc structure:
Ag$|$Fe(1L)$|$Vac(3L)$|$Ag(111) with total magnetic moment of
$2.81\,\mu_{B}$ per Fe atom. So, the maximum magnetic
moment density here is $39\,\mu_{B}/%
\operatorname{nm}%
^{2}$. The results for the mixing conductances is summarized in Fig.
\ref{local_picture}. We find that the ratio of $G^{\uparrow\downarrow}$ to the
(Ag) Sharvin conductances monotonically increases with the Fe density at the
Ag$|$Vac$\ $interface. The increase is linear at small densities and saturates
around $30\,\mu_{B}/%
\operatorname{nm}%
^{2}$ due to interactions between neighboring moments. We find that $%
\operatorname{Re}%
G^{\uparrow\downarrow}$ of Ag$|$YIG and Ag$|$Fe$|$vacuum agrees well for
corresponding Fe densities at the interface, in strong support of the local
moment model. 

Since the mixing conductance is dominated by the local moments at the
interface, we understand that the results are relatively stable against the
difficulties density functional theory has for insulators. The variation of
the band gap of the insulator as well as the band alignment with respect to
normal metal changes the penetration of the spin accumulation, but since only
the uppermost layers contribute this is of little consequence.

\begin{table}[h]
\caption{$G^{\uparrow\downarrow}$ of a disordered Ag$|$YIG(001) interface with
YFe-termination. Directional disorder is introduced by flipping three majority
Fe spins in the $2\times2$ super cell. }%
\label{disorder}
\begin{center}%
\begin{tabular*}
{8.5cm}[c]{@{\extracolsep{\fill}}ccc}\hline\hline
\  & $\operatorname{Re}G^{\uparrow\downarrow}\;(10^{14}\operatorname{\Omega
}^{-1}\operatorname{m}^{-2})$ & $\operatorname{Im}G^{\uparrow\downarrow
}\;(10^{14}\operatorname{\Omega}^{-1}\operatorname{m}^{-2})$\\\hline
clean & 3.010 & 0.302\\
disorder & 3.145 & 0.382\\\hline\hline
\end{tabular*}
\end{center}
\end{table}

Table \ref{disorder} shows the effect of directional disorder of magnetic
moments on the mixing conductance for Ag$|$YIG(001) with YFe-termination, for
which the integrated surface magnetic moment density is close to zero. Here,
we use a $2\times2$ lateral YIG supercell in which three magnetic moments are
flipped to a negative value, amounting to a total surface magnetic moment of
$-23.7\,\mu_{B}$ per lateral unit cell. The directional disorder of magnetic
moments at the interface slightly enhances $%
\operatorname{Re}%
G^{\uparrow\downarrow}$ (around $5\%$), as indeed expected from the local
moment picture.

\begin{figure}[h]
\includegraphics[width=8.6cm]{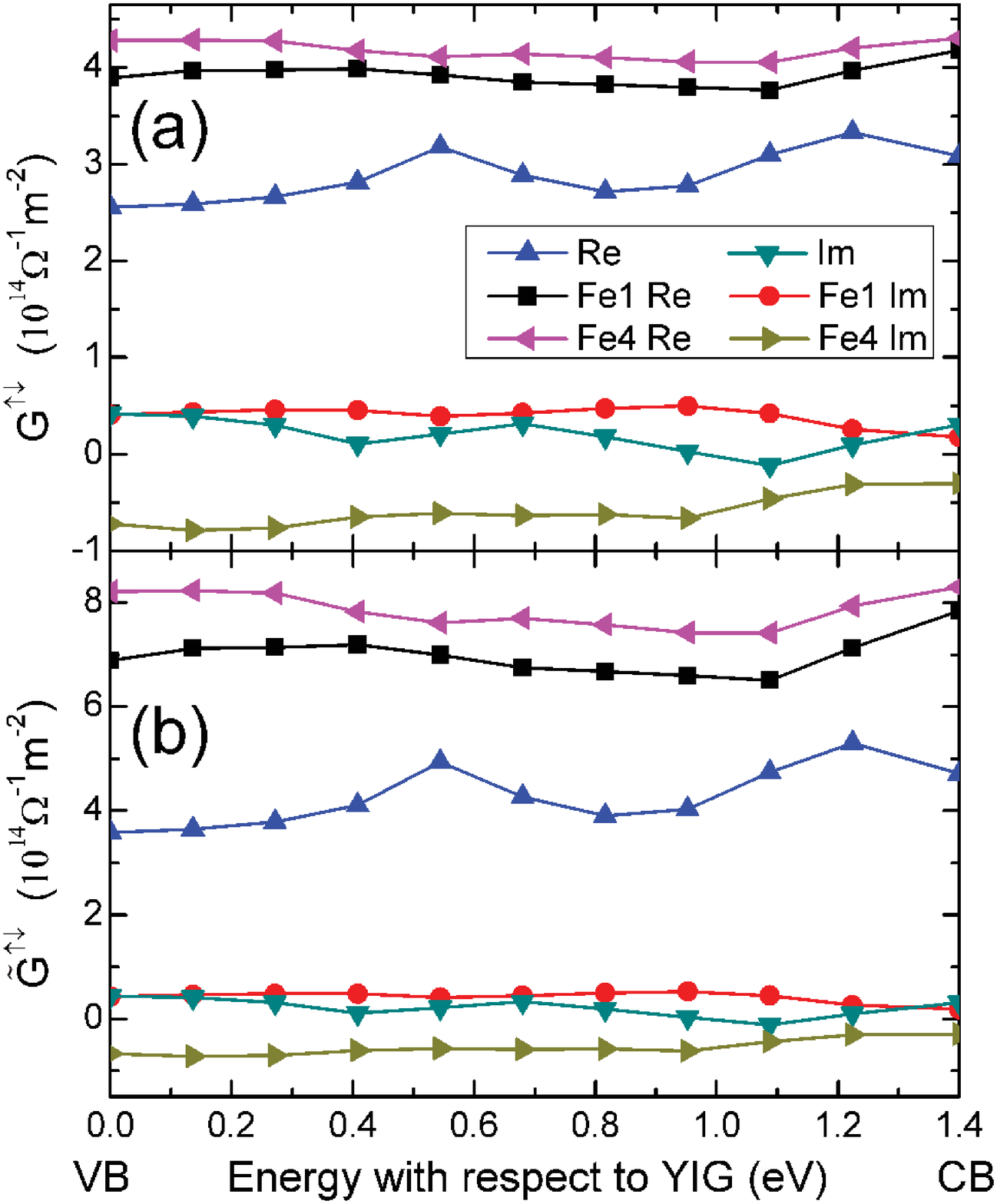}\newline
\caption{Spin mixing conductance of
Ag$|$Fe$_{n}|$YIG$|$Fe$_{n}|$Ag(001) with YFe termination for (a)
ballistic and (b) diffusive transport (i.e. in the presence of Schep
correction), where $n$ is the number for Fe monolayers inserted
between Ag$\ $and YIG. }
\label{fe_interpolation}%
\end{figure}

Inserting a thin ferromagnetic metallic layer between the normal
metal and YIG should enhance the spin mixing conductance. In Fig.
\ref{fe_interpolation} (a), we show that inserting Fe atomic layers
indeed
increases $%
\operatorname{Re}%
G^{\uparrow\downarrow}$ by 40-65\% up to the intermetallic Ag$|$Fe value,
which is close to the Ag Sharvin conductance.

In these calculation, the Ag reservoirs has been assumed to be
ballistic. When the spin mixing conductance is small relative to the
Sharvin conductance, this is a valid approximation, but otherwise
the diffusive nature of transport
may not be be neglected. Since $%
\operatorname{Re}%
G^{\uparrow\downarrow}$ turns out to be of the same order as
$G_{Ag}^{Sh}$ we have to introduce the diffusive transport
correction as introduced by Schep \textit{et al.} as
\cite{schep,schep02}
\begin{equation}
\frac{1}{\tilde{G}^{\uparrow\downarrow}}=\frac{1}{G^{\uparrow\downarrow}%
}-\frac{1}{2G_{Ag}^{Sh}}.
\end{equation}
The results are shown in Fig. \ref{fe_interpolation}(b). We observe that the
"Schep" correction enhances the spin mixing conductance by 20\% for for the
and about 90\% for 4 nonolayers Fe insertions between Ag$\ $and YIG.

The spin transfer can be maximized by a high density of magnetic ions at the
interfaces. In YIG we could not identify interface directions or cuts that are
especially promising, but this could be different for other magnetic
insulator, such as ferrites \cite{Slonczewski2010}. Slonczewski
\cite{Slonczewski2010} uses a local moment model with a somewhat smaller
exchange splitting (0.5$\,%
\operatorname{eV}%
$) than found here; when defined as $\triangle\left(  \vec{R}\right)
=\int_{\Omega_{WS}}\left(  V_{xc}^{\downarrow}\left(  \vec{R},\vec
{r}\right)  -V_{xc}^{\uparrow}\left( \vec{R},\vec{r}\right) \right)
\rho\left(  \vec{R},\vec{r}\right) d\vec{r},$ where $\rho\left(
\vec{R},\vec{r}\right)  $ is the density of the evanescent wave
function in YIG at mid-gap energy disregarding its spin splitting
\cite{Gunnarsson1976}, $\Omega_{WS}$ the Wigner-Seitz sphere at the
lattice site $\vec{R}$, and $V_{xc}^{\uparrow\left(
\downarrow\right)  }$ denotes the exchange-correlation potentials
for spin-up (down) electronsthe exchange splitting felt by the Ag
conduction electrons at the YIG interface is up to
$\sim3.0\,%
\operatorname{eV}%
$. Since Slonzcewski focusses on the magnetization dynamics of the magnetic
insulator we cannot carry out a quantitative comparison with his model here.

\section{Conclusion}

In conclusion, we computed the spin mixing conductance $G^{\uparrow\downarrow
}$ of the interface between silver and the insulating ferrimagnet Yttrium Iron
Garnet (YIG). $%
\operatorname{Re}%
G^{\uparrow\downarrow}$ is found to be of the order of 10$^{14}$ $%
\operatorname{\Omega }%
^{-1}%
\operatorname{m}%
^{-2}$, which is much larger than expected for a Stoner model, which indicates
the importance of the local magnetic exchange field at the interface. On the
other hand, $G^{\uparrow\downarrow}$ is not very sensitive to crystal
orientation and interface cut. $%
\operatorname{Re}%
G^{\uparrow\downarrow}$ can be enhanced to around 40-65\% of the fully
metallic limit by inserting a monolayers of iron between Ag and YIG. The
discrepancy between the measured and calculated mixing conductance might
indicate previously unidentified interface contaminations that, when removed,
would greatly improve the usefulness of magnetic insulators in spintronics.

\acknowledgments We would like to thank Burkard Hillebrands, Eiji Saitoh, and
Ken-ichi Uchida for stimulating discussions. This work was supported by
National Basic Research Program of China (973 Program) under the grant No.
2011CB921803 and NSF-China grant No. 60825404, the EC Contract ICT-257159
\textquotedblleft MACALO\textquotedblright\ and the Dutch FOM foundation. This
research was supported in part by the Project of Knowledge Innovation Program
(PKIP) of Chinese Academy of Sciences, Grant No. KJCX2.YW.W10.

\emph{Additional remark:} After first submission of our manuscript
arXiv:1103.3764, a manuscript was submitted and accepted by Physical
Review Letters (Heinrich B. et al., Phys. Rev. Lett., 107 (2011)
066604) that reports a mixing conductance that is clearly enhanced
compared to ref. [1], but still an order of magnitude smaller than
our predictions.

\end{document}